\documentstyle[osa,epsf,eqsecnum,epsfig,graphics,cite,elsart12]{revtex}

\def\lsim{\raise0.3ex\hbox{$<$\kern-0.75em\raise-1.1ex\hbox{$\sim$}}}
\def\gsim{\raise0.3ex\hbox{$>$\kern-0.75em\raise-1.1ex\hbox{$\sim$}}}

\title{Varying and inverting the mass hierarchy in collisional energy loss}
\author{Rodion Kolevatov$^{\rm 1}$ and Urs Achim Wiedemann$^{\rm 2}$}

\address{$^{\rm 1}$ Department of High Energy Physics, Saint-Petersburg State University, Saint-Petersburg, Ulyanovskaya, 1\\
$^{\rm 2}$ Department of Physics, CERN, Theory Division,
CH-1211 Gen\`eve 23, Switzerland}

\begin{document}

\maketitle

\begin{abstract}
Heavy ion collisions at RHIC and at the LHC give access to the medium-induced
suppression patterns of heavy-flavored single inclusive hadron
spectra at high transverse momentum. This opens novel opportunities for a detailed
characterization of the medium produced in the collision. In this note, we point out that the
capacity of a QCD medium to absorb the recoil of a partonic projectile is
an independent signature, which
may differ for different media at the same density. In particular, while the mass hierarchy
(i.e., the projectile mass dependence) of radiative energy loss depends solely on a property
of the projectile, the mass hierarchy of collisional energy loss depends significantly on 
properties of the medium. By varying these properties in a class of models, we find that
the mass hierarchy of collisional parton energy loss can be modified considerably and
can even be inverted,
compared to that of radiative parton energy loss. This may help to disentangle
the relative strengths of radiative and collisional contributions to jet quenching, and it
may be employed to constrain properties of the produced QCD medium beyond its density. 
\end{abstract}

\section{Introduction}
'Jet quenching', the energy loss of high momentum partons in hot and dense QCD matter
is at the basis of the strong medium-induced suppression of hadronic high-$p_T$ spectra, 
discovered at RHIC~\cite{RHIC}. It is also expected to dominate the physics at high 
transverse momenta in heavy ion collisions at the LHC~\cite{LHC}. The strong
sensitivity of high-$p_T$ hadronic spectra to the nuclear environment makes jet quenching 
a promising tool for the characterization of properties of the matter produced in heavy ion 
collisions. However, the accuracy of this tool depends 
largely on understanding the microscopic mechanism by which the medium affects the 
stopping and fragmentation of partonic projectiles. At sufficiently high projectile energy, 
an essentially recoilless radiative energy loss mechanism is expected to dominate on
general kinematic grounds~\cite{Baier:1996sk,Zakharov:1997uu,Wiedemann:2000za,Gyulassy:2000er,Wang:2001if}. At sufficiently small projectile energies, 
however, recoil is expected to be non-negligible. Several recent model 
studies~\cite{Djordjevic:2006tw,Wicks:2005gt,Wicks:2007zz,Adil:2006ei,Peigne:2008nd}
attribute a sizable role to collisional mechanisms mediated via elastic interactions.
But an experimental strategy
to disentangle the effects of elastic and inelastic interactions (a.k.a. collisional and radiative
energy loss) is missing so far. This makes it interesting to look for experimental signatures
which are qualitatively different for both mechanisms.

For the radiative energy loss mechanism, we know that the dependence of parton energy 
loss on the mass and color charge of the partonic projectile shows a characteristic hierarchy.  
Due to their larger color charge, gluons radiate more than light quarks. 
And light quarks radiate more than heavy quarks, since radiation is suppressed as a function
of projectile 
mass~\cite{Dokshitzer:2001zm,Armesto:2003jh,Zhang:2003wk,Djordjevic:2003zk,Armesto:2005iq}.
Remarkably, this hierarchy of radiative parton energy loss is determined 
solely by properties of the partonic projectile, namely its color charge and its mass. 
Properties of the medium affect the absolute
strength of medium modifications, but leave its relative dependence on parton identity
unchanged. The current understanding of the characteristic
tell-tale signs of collisional energy loss is less complete. To the best of our knowledge, 
one has not yet addressed the question to what extent the mass hierarchy of collisional 
energy loss depends on properties of the medium or on properties of the projectile. 

To discuss medium modifications of parton propagation, one must specify the properties
of the medium. If the medium is in thermal equilibrium, then all properties of the medium depend 
on temperature only. In principle, this fixes the relative strength of collisional and radiative 
energy loss. In practice, however, this relative strength may be difficult to evaluate for the
temperature range which is in reach of heavy ion collisions. In addition, the systems 
produced in heavy ion collisions may show interesting deviations from the idealization of
a thermal heat bath. For these reasons, we do not presuppose in the following a unique 
relation between the density of the medium and other features (such as the masses
of quasi-particles or the capacity to absorb recoil). Rather, we shall vary these features
 independently within some parameter range with the view of determining them finally
in a comparison to data. 

In section~\ref{sec2} of this paper, we introduce a class of models of the medium, and we 
discuss how these models differ in their capacity of absorbing recoil. 
In section ~\ref{sec3} and \ref{sec4}, we then point out that depending on the model-dependent 
capacity of the medium to absorb recoil, the strength of collisional energy loss and its
dependence on projectile mass can vary strongly. 
 We finally comment on the implications of our findings. 

\section{The model}
\label{sec2}

Several calculations of radiative  energy loss model the medium in terms of a set of static colored scattering centers~\cite{Baier:1996sk,Zakharov:1997uu,Wiedemann:2000za,Gyulassy:2000er}. 
Since medium-induced radiation is expected to depend mainly on the transverse color
field strength presented by the medium to the projectile, and since a set of static scattering
centers parametrizes conveniently this color field strength, such a simple model
captures the main feature relevant for radiative energy loss. In the same spirit, many recent 
calculations of collisional energy loss model the medium as either a set of 
massless particles with thermal momentum distribution~\cite{Djordjevic:2006tw,Wicks:2005gt}, 
or as a set of initially static massive scattering centers~\cite{Wicks:2007zz}. By making
the target scattering centers dynamical, these models parametrize not only the color
field strength but also the capacity of the medium to absorb recoil. 

Following this approach, we consider models, which are characterized by two
parameters: the mass $m_t$ of colored scattering centers in the fundamental ($q$) and 
adjoint ($g$) representation, and a parameter $T$ which characterizes the momentum 
distribution of these scattering centers,
\begin{equation}
 n_q(k)=f_{\rm scale}\, \frac{1}{(2\pi)^3}\frac{12 n_f}{e^{E_k/T}+1}\, , \quad 
 n_g(k)=f_{\rm scale}\, \frac{1}{(2\pi^3)}\frac{16}{e^{E_k/T}-1}\, ,
 \label{distr}
\end{equation}
where $E_k = \sqrt{m_t^2 + k^2}$ and $n_f=2$. The mass $m_t$ does not depend
on the temperature $T$. If the scale factor is $f_{\rm scale} = 1$, then the number
density of scattering centers decreases with increasing $m_t$. To disentangle
observable consequences of a decreasing density from other target mass dependent
effects, we also consider the case
\begin{equation}
 f_{\rm scale}(m_t) = \frac{\int d^3k\, (n_q(k) + \frac{9}{4} n_g(k)) \vert_{m_t = 200\, {\rm MeV}} }{
 		\int d^3k\, (n_q(k) + \frac{9}{4} n_g(k)) \vert_{m_t} }\, .
		\label{fscale}
\end{equation}
With this normalization, the integrated effective density of scattering centers entering parton 
energy loss (see eq. (\ref{eloss}) below) does not depend on $m_t$. 

We consider a light or heavy projectile quark $Q$, which propagates through the 
class of model targets described above. 
This quark accumulates collisional momentum loss $\Delta p_Q$ per unit path length $dx$ by 
incoherent elastic scattering on target partons,
%
%
\begin{equation}
\frac{d\Delta p_Q}{dx}= \frac{1}{v_Q} \int dp_f (p-p_f) \int k^2 dk (n_q(k) + \frac{9}{4} n_g(k)) 
\frac{d\sigma^{\rm int}_{Qq}(k,p_f)}{dp_f}\, . \label{eloss}
\end{equation}
Here, $p$ is the initial and $p_f$ the final momentum of  the projectile $Q$, and 
$v_Q$ denotes its velocity in the rest frame of the medium. We
approximate the elastic $Q\, g$ scattering cross section 
by the leading $t$-channel exchange, such that 
$\frac{d\sigma^{\rm int}_{Qg}(\langle k \rangle,p_f)}{dp_f} =
\frac{C_A}{C_F}\frac{d\sigma^{\rm int}_{Qq}(\langle k \rangle,p_f)}{dp_f}$. The
elastic scattering cross section is of the general form
\begin{equation}
\frac{d\sigma^{\rm int}}{dp_f}=2\pi \int {\rm d}(\cos \psi) \frac{1}{4p^0 k^0} |{\cal{M}}|^2 d\Phi\, ,
\label{eq2}
\end{equation}
where $2\pi\int {\rm d}(\cos \psi)$ denotes the integration over the direction of the incoming
target particle, and $d\Phi$ denotes the phase space volume.
The model is fully defined once the scattering matrix is specified.

For the elastic scattering matrix element ${\cal M}$, we use the expression to lowest order in 
$\alpha_s$ with single gluon exchange in the t-chanel described by the HTL-resummed 
propagator \cite{Kalashnikov:1979cy,Klimov:1982bv}. This is the starting point of many works on collisional energy loss \cite{Thoma:1990fm,Braaten:1991we}. 
As in the recent work of Djordjevic~\cite{Djordjevic:2006tw}, we compute the matrix 
element for Q-q scattering without any assumption on the smallness of masses or energy 
transfers. In summary, our calculation is a standard calculation of collisional energy loss 
in which $t$-channel exchanges are regulated by thermal
propagators, but in which - in contrast to previous models - the mass of the scattering
centers in the medium is treated as an independent parameter $m_t$. 
Also, in contrast to the constant coupling constant $\alpha_s = 0.3$ used 
in~\cite{Djordjevic:2006tw,Wicks:2005gt}, we use a running coupling constant
$\alpha_s(\mu_D^2 + k_T^2)$, where $\mu_D^2 = g(\mu_D^2)^2\, T^2\, \left(1 + n_f/6\right)$ 
defines the Debye screening mass~\cite{Peshier:2006ah}.  
For $T=225$~MeV, this choice corresponds to $\mu_D = 680$~MeV.

%
\begin{figure}[t]
 \centering
\includegraphics[width=10cm]{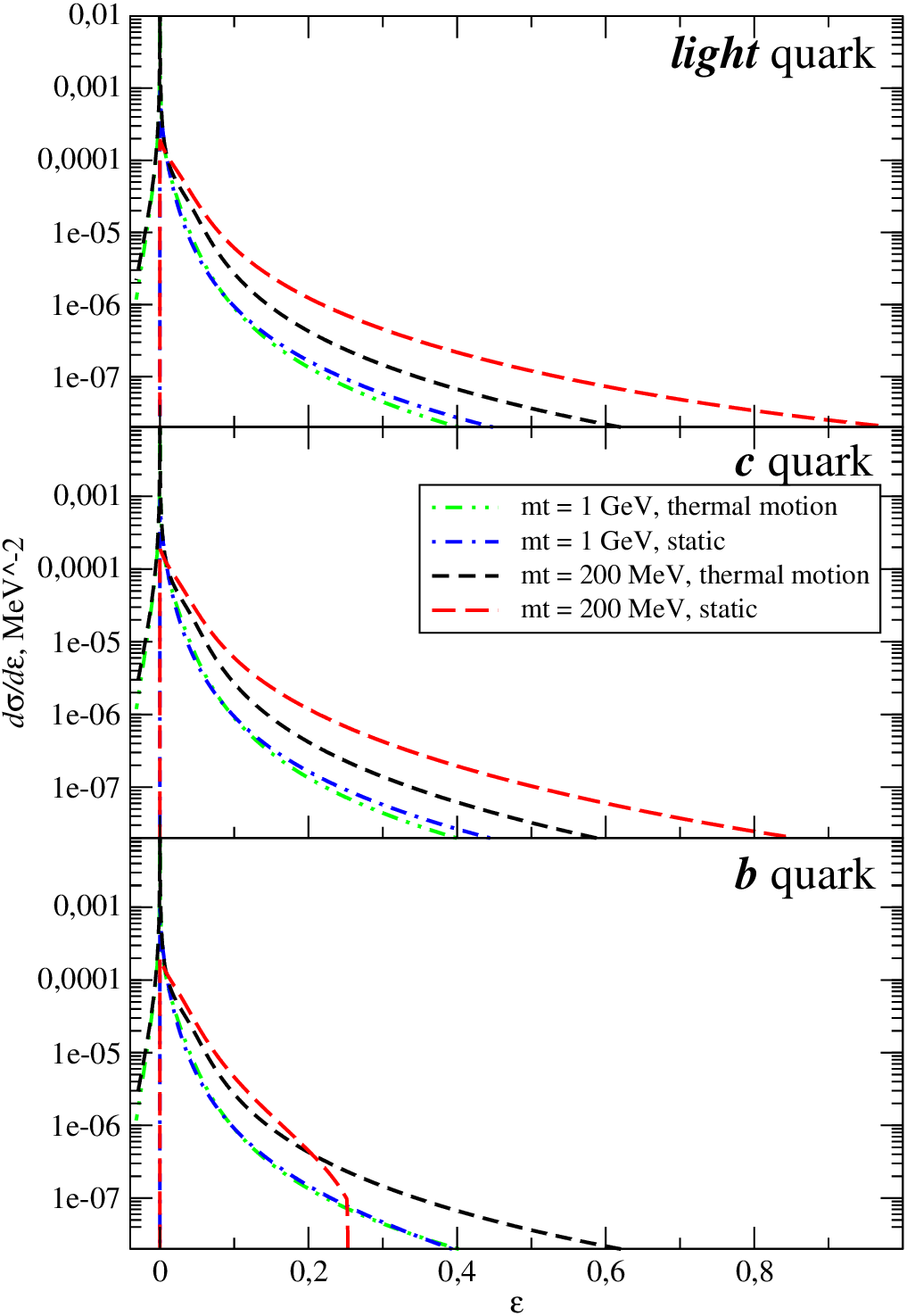}
\caption{The cross section of fractional energy loss $\epsilon = \Delta E/E$ 
in one collision, calculated from (\ref{eq2}) for different partonic projectiles and 
for different masses $m_t$
of the scattering centers in the medium. The initial projectile momentum is $p = 20$ GeV.
The target scattering centers are either initially at rest ('static'), or they show a thermal
momentum distribution with $T=225$~MeV ('thermal motion'). In the latter case, the cross section is averaged over all possible target particle momenta.}
\label{fig1}
\end{figure}
\section{Fractional collisional energy loss}
\label{sec3}
In Fig.~\ref{fig1}, we plot the cross section for the fractional energy loss $\epsilon = \Delta E/E$,
suffered by a relativistic $p=20$ GeV partonic projectile in a single collision. As partonic 
projectiles, we consider light quarks ($m_q=200$~MeV), charm quarks ($m_c=1200$~MeV) and 
bottom quarks ($m_b=4750$~MeV). The scattering centers in the medium have a target mass 
$m_t$.  In general, one sees from Fig.~\ref{fig1} that collisional
energy loss peaks always at relatively small momentum fractions. This comes from the
fact, that elastic interactions are dominated by small-angle scattering, in which longitudinal
momentum transfer is small.

Fig.~\ref{fig1} compares the fractional energy loss for a model, in which the scattering centers 
in the target are initially at rest ('static'), to a model in which their distribution is thermal
 with $T = 225$ MeV. In heavy ion collisions, one expects that the scattering centers in 
 the medium show some random (thermal) motion, so one favors a thermal distribution
 on physical grounds. But static distributions have been considered recently~\cite{Wicks:2007zz} 
for scattering centers with target masses $m_t = 200$ MeV. Our results for this case
agree with those of Ref.~\cite{Wicks:2007zz}. This illustrates that our set-up is consistent
with standard collisional energy loss calculation.  An increase
of the target mass  from $m_t = 200$ MeV to $m_t = 1$ GeV in this static case leads 
generally to a significant reduction of collisional energy loss. In this sense, the capacity
of this model of the medium to absorb recoil can be varied by varying $m_t$. This is,
of course, expected on general kinematic grounds. In the limit of infinitely 
massive scattering centers in the target, the medium would not absorb any 
recoil. The collisional interactions would not be visible
in an energy degradation of the projectile, but only in its momentum broadening.

The maximal possible value of momentum transfer in a single collision is restricted
by the available phase space. This restriction is especially pronounced for an initially 
static target. For instance, a heavy projectile, colliding with a light static target cannot
loose but a certain fraction of its total energy. As is seen in the lower panel of Fig.~\ref{fig1}, 
this leads for heavy $b$-quark projectiles to a severe restriction of energy loss for light 
target mass. The restriction is weakened if the target particle 
is heavier or if it carries some initial randomly distributed momentum. 

In comparing the models of scattering centers with thermal and with static momentum distribution,
shown in Fig.~\ref{fig1}, we note that collisional energy loss for a relativistic projectile
appears to depend mainly
on the average total energy of the scattering centers in the target,
rather than on their mass. Since a light particle in contact with a heat bath 
of temperature $T=225$~MeV has an average kinetic momentum of $\gsim 600$ MeV, 
this explains qualitatively the significant differences between both models for a small target 
mass $m_t = 200$ MeV, where the average total energy of scattering centers exceeds their 
rest mass by far. The same argument accounts for the much less pronounced
differences for $m_t = 1$ GeV. It may also explain why the case of small target mass 
$m_t=200$ MeV with thermal motion lies in between the curves for static scenarios
with $m_t = 200$ MeV and $m_t = 1$ GeV, respectively. 

A thermal distribution of scattering centers leads also to some novel
features in the cross section of fractional energy loss. In particular, the cross sections
in Fig.~\ref{fig1} show non-vanishing contributions for negative $\epsilon$, since a
projectile may (though with small probability) gain energy in scattering with a target 
component. With increasing target mass, the probability of loosing a significant 
fraction of the total initial energy decreases for the same generic kinematic reasons
as in the case of the static scenario, described above. So, also in this case varying the
target mass changes the capacity of the medium to absorb recoil.

\section{Mass Hierarchy of Collisional Energy Loss}
\label{sec4}

\begin{figure}[t]
 \centering
\includegraphics[width=14cm]{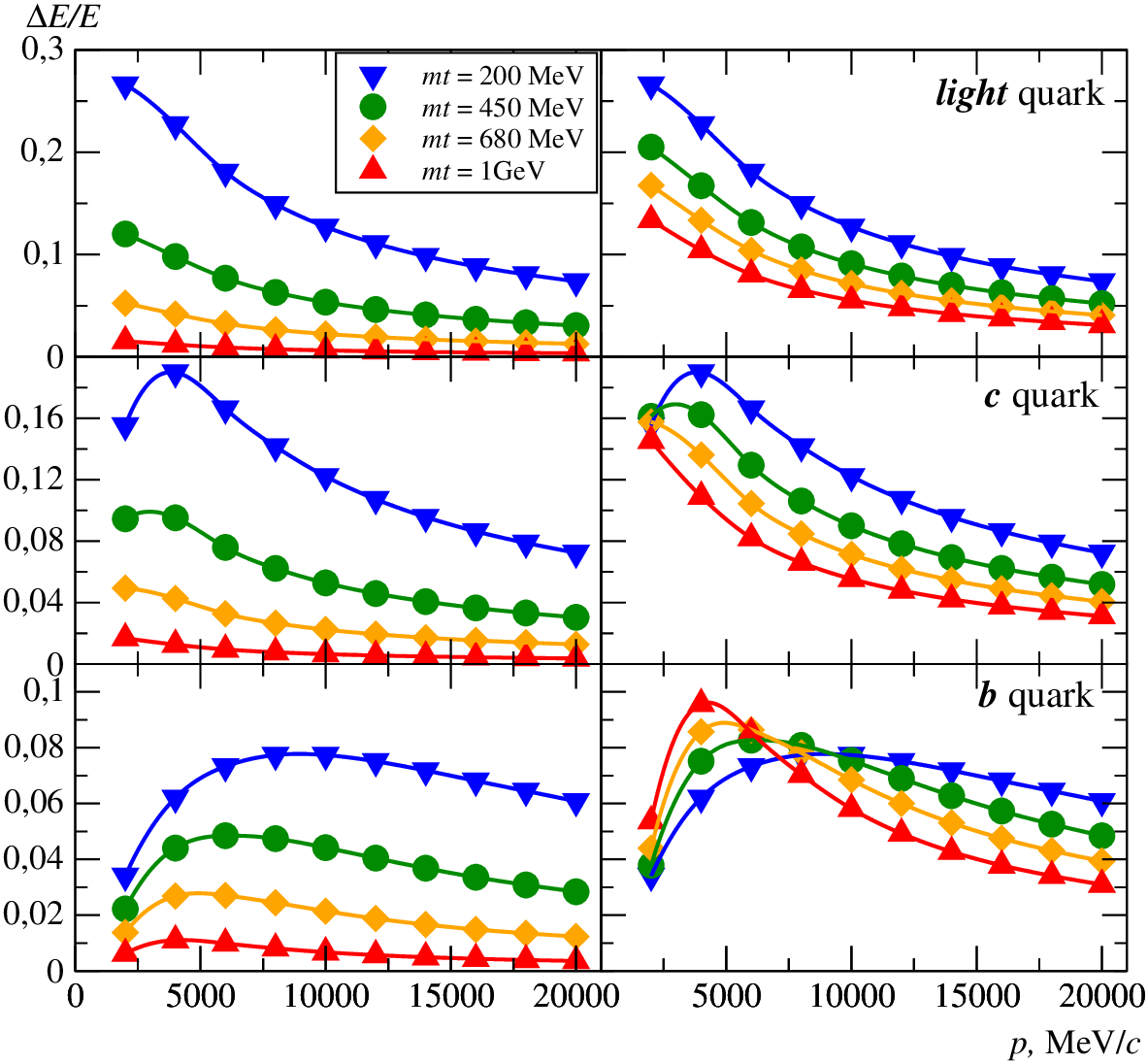}
\caption{Collisional parton energy loss fraction $\Delta E/E$ as a function of projectile
momentum $p$ calculated from equation (\ref{eloss}) for light quarks ($m_q = 200$~MeV, 
upper panel), charm quarks ($m_c = 1.2$~GeV, middle panel) and bottom quarks 
($m_b = 4.75$~GeV, lower panel). The path length $L=5$~fm, other parameters are 
chosen as for Fig.~\ref{fig1}. The massive target particles in the medium are 
distributed according to (\ref{distr}) with $f_{\rm scale} = 1$ (left column) or
$f_{\rm scale}$ defined by eq. (\ref{fscale}) (right column). } 
\label{fig2}
\end{figure}

Fig.~\ref{fig2} shows results for the average energy loss of a partonic
projectile of momentum $p$, which propagates through $L=5$ fm of matter. 
The medium is characterized by scattering centers of target mass $m_t$. We consider
{\it i)} the case for which the massive target particles show a thermal distribution 
[eq. (\ref{distr}) with $f_{\rm scale} =1$] and {\it ii)} the case that the momentum
distribution of target particles follows case {\it i)} but that the total density is fixed
by eq. (\ref{fscale}) to an $m_t$-independent value. For projectile particles whose momentum
does not differ much from typical momenta in the media, the distinction between
projectile and target becomes questionable. For this reason, we show results
for projectile momenta of $p > 2$ GeV only, though our model would smoothly extend to
softer momenta. 

 We consider first the case {\it i)} that the massive target particles follow the thermal
 distribution (\ref{distr}) with $f_{\rm scale} =1$. In this case, the average collisional 
 energy loss drops strongly with increasing target mass for all values of projectile
 momentum, see left hand side of Fig.~\ref{fig2}. For $m_t = 1$ GeV, the average 
 energy loss is a factor of order 10 smaller than for $m_t = 200$ MeV. This 
 strong dependence on target mass $m_t$ has two different origins. First, by increasing 
 $m_t$, the density of scattering centers decreases and this leads to a strong decrease of 
 $\Delta E$. Second, changing $m_t$ also changes the average energy loss per collision.
 
\begin{figure}[!ht]
 \centering
\includegraphics[width=10cm]{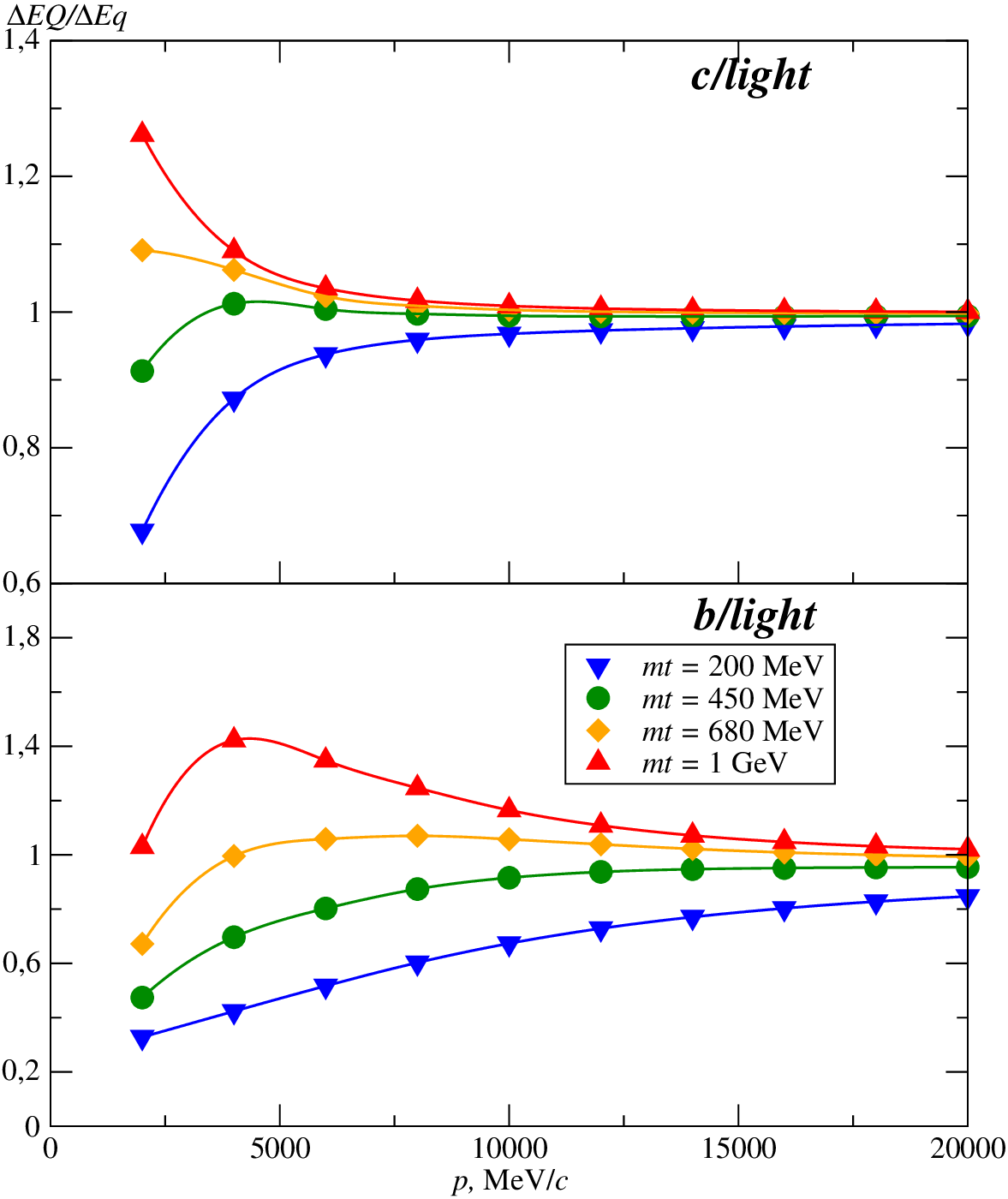}
\caption{The heavy-to-light ratio $\Delta E_Q / \Delta E_q$ of collisional energy loss for 
charm quarks (upper panel) and bottom quarks (lower panel), compared to that of 
light quarks ($m_q = 200$~MeV). The results for the numerator $\Delta E_Q$ and the 
denominator $\Delta E_q$ are the same as used for plotting Fig.~\ref{fig2}. }
\label{fig3}
\end{figure}
 
 By rescaling the density of scattering centers with equation (\ref{fscale})
 [This case {\it ii)} is shown on the right hand side of Fig.~\ref{fig2}.], we consider a 
 class of model media in which the total number density of scattering centers is independent
 of $m_t$. This eliminates the first source of the strong $m_t$-dependence. So, these
 media have the same entropy density, irrespective of $m_t$, but they differ in their capacity 
 to absorb recoil.
 At sufficiently high projectile momentum $p$, we see that the capacity to absorb recoil
 decreases with increasing target mass $m_t$. However, for sufficiently small momenta, 
 when the mass of the projectile is not negligible for the scattering process, 
 heavier projectiles can transfer their  energy more efficiently to more massive 
 scattering centers, than to light ones. This leads to an inversion of the $m_t$-dependence
 of the average energy loss, which is particularly pronounced for the bottom quark
 (see lower right panel in Fig.~\ref{fig2}).

One expects that for large projectile momentum, 
collisional energy loss depends negligibly on projectile mass. Consistent with this expectation,
the curves for $\Delta E/E$ for fixed $m_t$ but different projectile mass approach
the same numerical value in the limit of large projectile momentum $p$. This is more
clearly seen in Fig.~\ref{fig3}, where the ratio $\Delta E_Q/ \Delta E_q$ approaches
unity for large projectile momentum $p$. 

For finite projectile momentum $p$, Fig.~\ref{fig3} shows that a property of the medium
(namely the value of $m_t$) determines whether the collisional energy loss of a massive 
partonic projectile is larger or smaller than that of a light projectile. This is in stark contrast 
to the case of radiative energy loss, where the dependence on projectile mass is expected 
to be governed entirely by the deadcone effect~\cite{Dokshitzer:2001zm}, which depends 
on properties of the projectile 
only. Some rough qualitative aspects of the inversion of the projectile mass hierarchy, seen
in Fig.~\ref{fig3}, may be understood on kinematical grounds. If the projectile is much 
heavier than the target scattering center, it has the tendency to simply 'run over'
the target without significant change of momentum. In comparison, a light
projectile of same momentum can transfer a larger fraction of its momentum to a
target of similar mass. This implies
that for a light target, the ratio $\Delta E_Q/\Delta E_q$ is smaller than unity. 
As the target mass is increased, the collisional energy loss of light 
projectiles always decreases, as seen in Fig.~\ref{fig2}. However, if the total density of 
scattering centers is kept fixed, the energy loss of heavy quarks can increase with increasing
$m_t$ for the physics reasons explained in the context of Fig.~\ref{fig2} above. This 
implies that for sufficiently large $m_t$, $\Delta E_Q/\Delta E_q$  grows above unity:
the projectile mass hierarchy is inverted.

\section{Discussion}
It has been pointed out recently that collisional parton energy loss can contribute to a
medium-induced energy degradation of highly energetic partons in heavy ion collisions,
which is comparable for heavy and for light quarks. This is in contrast to radiative
energy loss mechanisms, which degrade the energy of massive projectiles less 
efficiently. The present study shows that, if suitable assumptions about the
recoil properties of the medium are made, then the energy of heavy quarks may be quenched
indeed as efficiently as that of light quarks. However, this finding depends strongly on the
model-dependent properties of the medium. With relatively small changes of the 
model parameter $m_t$, one can also realize scenarios, in which light quarks are suppressed
twice as much as bottom quarks for $p< 10$~GeV, or in which bottom quarks are suppressed
more than light quarks. More generally, in the model scenarios explored here,
small target masses are preferred if one wants to obtain a sizable contribution from collisional energy 
loss, but relatively large target masses are needed to obtain similar suppression factors
for light and heavy quarks. A phenomenologically successful modeling of similar
nuclear modification factors for light- and heavy-flavored hadrons in terms of collisional
energy loss will need to satisfy these complementary constraints. We conclude that, if the 
relative strength of collisional energy loss can be determined independently, 
then the strong sensitivity of collisional energy loss on $m_t$ provides information
about a property of the medium, which is distinct from its density.

We finally remark that determining a value of $m_t$ in a model comparison to 
data does not imply a fortiori that the medium is a gas of quasi-particle of mass $m_t$.
Rather, such an agreement could also be consistent with the picture of a strongly
coupled medium, that does not carry quasi-particle excitations, but that absorbs
recoil at a rate comparable to a gas of quasi-particles of mass $m_t$. In this latter
case, which has received considerable support from RHIC data~\cite{RHIC}, the 
models explored here may still be able to parametrize with $m_t$ the magnitude of
longitudinal momentum transfer from the projectile to the medium, but they would
be inadequate for a description of the strongly coupled dynamics of the medium.

{\bf Acknowledgements} This work was partially supported by INTAS
05-112-5031, grant RNP.2.2.2.2.1547, RFBR-CERN Nr~08-02-91004-CERN\_a
and NFR Project 185664/V30. We thank Grigory Feofilov for discussion and
support throughout this work. We also acknowledge discussions with 
Andrey Tarasov.



\begin{thebibliography}{99}

%
\bibitem{RHIC}
  K.~Adcox {\it et al.}  [PHENIX Collaboration],
  Nucl.\ Phys.\ A {\bf 757} (2005) 184.
%
  B.~B.~Back {\it et al.} [PHOBOS Collaboration],
  Nucl.\ Phys.\ A {\bf 757} (2005) 28.
%
  I.~Arsene {\it et al.}  [BRAHMS Collaboration],
  Nucl.\ Phys.\ A {\bf 757} (2005) 1.
%
  J.~Adams {\it et al.}  [STAR Collaboration],
  Nucl.\ Phys.\ A {\bf 757} (2005) 102.

\bibitem{LHC}
  F.~Carminati {\it et al.}  [ALICE Collaboration],
  J.\ Phys.\ G {\bf 30} (2004) 1517.
 %
  B.~Alessandro {\it et al.}  [ALICE Collaboration],
  J.\ Phys.\ G {\bf 32} (2006) 1295.
%
  D.~d'Enterria {\it et al.} [CMS Collaboration],
  J.\ Phys.\ G {\bf 34} (2007) 2307.

\bibitem{Baier:1996sk}
R.~Baier, Y.~L.~Dokshitzer, A.~H.~Mueller, S.~Peign\'e and D.~Schiff,
Nucl.\ Phys.\ B {\bf 484} (1997) 265.
%
 \bibitem{Zakharov:1997uu}
B.~G.~Zakharov,
JETP Lett.\  {\bf 65} (1997) 615.
%
\bibitem{Wiedemann:2000za}
U.~A.~Wiedemann,
Nucl.\ Phys.\ B {\bf 588} (2000) 303.
%
\bibitem{Gyulassy:2000er}
M.~Gyulassy, P.~Levai and I.~Vitev,
Nucl.\ Phys.\ B {\bf 594} (2001) 371.
%
\bibitem{Wang:2001if}
X.~N.~Wang and X.~F.~Guo,
Nucl.\ Phys.\ A {\bf 696} (2001) 788.

\bibitem{Djordjevic:2006tw}
  M.~Djordjevic,
  Phys.\ Rev.\  C {\bf 74} (2006) 064907
  [arXiv:nucl-th/0603066].

\bibitem{Wicks:2005gt}
  S.~Wicks, W.~Horowitz, M.~Djordjevic and M.~Gyulassy,
  Nucl.\ Phys.\  A {\bf 784} (2007) 426
  [arXiv:nucl-th/0512076].

\bibitem{Wicks:2007zz}
  S.~Wicks and M.~Gyulassy,
  J.\ Phys.\ G {\bf 34} (2007) S989
  [arXiv:nucl-th/0701088].

\bibitem{Adil:2006ei}
  A.~Adil, M.~Gyulassy, W.~A.~Horowitz and S.~Wicks,
  Phys.\ Rev.\  C {\bf 75} (2007) 044906
  [arXiv:nucl-th/0606010].

\bibitem{Peigne:2008nd}
  S.~Peign\'e and A.~Peshier,
  arXiv:0802.4364 [hep-ph].

\bibitem{Dokshitzer:2001zm}
Y.~L.~Dokshitzer and D.~E.~Kharzeev,
Phys.\ Lett.\ B {\bf 519}  (2001) 199.
%
\bibitem{Armesto:2003jh}
N.~Armesto, C.~A.~Salgado and U.~A.~Wiedemann,
Phys. Rev. D {\bf 69} (2004) 114003.
%
\bibitem{Zhang:2003wk}
B.~W.~Zhang, E.~Wang and X.~N.~Wang,
Phys.\ Rev.\ Lett.\  {\bf 93} (2004) 072301.
%
\bibitem{Djordjevic:2003zk}
M.~Djordjevic and M.~Gyulassy,
Nucl.\ Phys.\ A {\bf 733} (2004) 265.

\bibitem{Armesto:2005iq}
  N.~Armesto, A.~Dainese, C.~A.~Salgado and U.~A.~Wiedemann,
  Phys.\ Rev.\  D {\bf 71} (2005) 054027
  [arXiv:hep-ph/0501225].



\bibitem{Kalashnikov:1979cy}
  O.~K.~Kalashnikov and V.~V.~Klimov,
  Sov.\ J.\ Nucl.\ Phys.\  {\bf 31}, 699 (1980)
  [Yad.\ Fiz.\  {\bf 31}, 1357 (1980)].

\bibitem{Klimov:1982bv}
  V.~V.~Klimov,
  Sov.\ Phys.\ JETP {\bf 55}, 199 (1982)
  [Zh.\ Eksp.\ Teor.\ Fiz.\  {\bf 82}, 336 (1982)].

%
\bibitem{Thoma:1990fm}
  M.~H.~Thoma and M.~Gyulassy,
  Nucl.\ Phys.\  B {\bf 351}, 491 (1991).
%
\bibitem{Braaten:1991we}
  E.~Braaten and M.~H.~Thoma,
  Phys.\ Rev.\  D {\bf 44}, 2625 (1991).

\bibitem{Peshier:2006ah}
  A.~Peshier,
  arXiv:hep-ph/0601119.
\end{thebibliography}
\end{document}